%% file: main.tex
\documentclass[conference]{IEEEtran}
\IEEEoverridecommandlockouts
\usepackage{svg}
\usepackage{cite}
\usepackage{amsthm}
\usepackage{amsmath,amssymb,amsfonts}
\usepackage{algorithmic}
\usepackage{graphicx}
\usepackage{textcomp}
\usepackage{xcolor}
\usepackage{bm}
\usepackage{caption}
\usepackage{subcaption}
\usepackage{multirow}
\usepackage{multicol}
\usepackage{caption}
\usepackage{subcaption}
\usepackage{geometry}
\begin{document}

\title{\Large\bf TAFA: Design Automation of Analog Mixed-Signal FIR Filters Using Time Approximation Architecture}

\author{\IEEEauthorblockN{Shiyu Su, Qiaochu Zhang, Juzheng Liu, Mohsen Hassanpourghadi, Rezwan Rasul, and Mike Shuo-Wei Chen}
\IEEEauthorblockA{\textit{Ming Hsieh Department of Electrical and Computer Engineering}\\
\textit{University of Southern California}, 
Los Angeles, CA 90089 \\
\{shiyusu, qiaochuz, juzhengl, mhassanp, rrasul, swchen\}@usc.edu}
}

\maketitle

\makeatletter
\def\ps@IEEEtitlepagestyle{%
  \def\@oddfoot{\mycopyrightnotice}%
  \def\@evenfoot{}%
}
\makeatother
\def\mycopyrightnotice{%
  \begin{minipage}{\textwidth}
    \footnotesize
    ~ \hfill\\~\\
  \end{minipage}
  \gdef\mycopyrightnotice{}
}

{\small\bf Abstract---
A digital finite impulse response (FIR) filter design is fully synthesizable, thanks to the mature CAD support of digital circuitry. On the contrary, analog mixed-signal (AMS) filter design is mostly a manual process, including architecture selection, schematic design, and layout. This work presents a systematic design methodology to automate AMS FIR filter design using a time approximation architecture without any tunable passive component, such as switched capacitor or resistor. It not only enhances the flexibility of the filter but also facilitates design automation with reduced analog complexity. The proposed design flow features a hybrid approximation scheme that automatically optimize the filter's impulse response in light of time quantization effects, which shows significant performance improvement with minimum designer's efforts in the loop. Additionally, a layout-aware regression model based on an artificial neural network (ANN), in combination with gradient-based search algorithm, is used to automate and expedite the filter design. With the proposed framework, we demonstrate rapid synthesis of AMS FIR filters in 65nm process from specification to layout.}


\hyphenation{op-tical net-works semi-conduc-tor}

\IEEEpeerreviewmaketitle

\section{Introduction}
High-performance filtering is increasingly demanded by wireless communication systems for supporting multi-band and multi-mode operations. Traditionally, bulky off-chip SAW filters or tuned LC filters are used at the transmitter and receiver front-ends to achieve the required selectivity. As a more scaling-friendly alternative of these conventional analog filter, an analog mixed-signal (AMS) finite impulse response (FIR) filter is gaining growing popularity in direct radio frequency (RF) transmitters \cite{bhat2017wideband,gaber2011cmos}. However, the reconfiguration of such filter is typically achieved by tuning the analog components, i.e., resistor, capacitor or current source. This leads to limited flexibility, makes the system vulnerable to process, voltage and temperature (PVT) variations, and imposes challenges for a complete design automation. 

Recently, a time-approximation filter (TAF) technique is proposed to tackle the issue by approximating the amplitude-varying impulse response of an AMS FIR filter with a constant-amplitude but time-duration-varying (i.e. digital-like) waveform \cite{su2020jssc}. Thanks to the mostly digital operations, TAF is highly flexible, scalable, implementation-friendly, and less prone to PVT variations. As a tradeoff of using time approximation, the finest achievable time resolution of TAF is constrained by the system clock rate, which degrades the filter performance and hence necessitates an effective searching algorithm to fine tune the filter's impulse response. Additionally, like most AMS circuit blocks, AMS filter design is mostly a manual process compared to a digital FIR filter. Although there are prior arts exploring digital architectures \cite{xu2017scaling, deng2014fully, zhang2021mdll} or using digital gates to approximate the function of an analog block \cite{chen2019digital, Liu2015cicc}, most of the works lack efficient modeling and optimization techniques for the sizing and biasing of the circuits, which still require significant amounts of handcrafted design efforts and time from experienced analog designers.

To address the existing design automation challenges for AMS filters, we propose a design automation flow, hereafter referred to as TAFA, with the following objectives: 1) apply time approximation technique to the AMS FIR filter and derive an optimum impulse response for the filter with minimum human intervention, and 2) automate the design of high-performance AMS filters within a short time frame. The primary features of TAFA are summarized below.

\begin{itemize}
    \item The proposed hybrid approximation scheme for filter's impulse response significantly reduces time approximation errors of TAF over a wide range of specifications.
    \item A mostly-digital capacitor digital-to-analog converter (DAC) structure with a counter-based TAF waveform generation is proposed for filter implementation which favors the technology scaling and digital design flow.  
    \item A layout-aware regression model combined with advanced search algorithms is used to expedite the synthesis of the filter by orders of magnitude during both schematic and layout design stages.
\end{itemize}

\section{Time-Approximation Filter}

\subsection{Time Approximation of an Arbitrary AMS FIR Filter}

\begin{figure}[!t]
    \centering
    \includegraphics[width=0.36\textwidth]{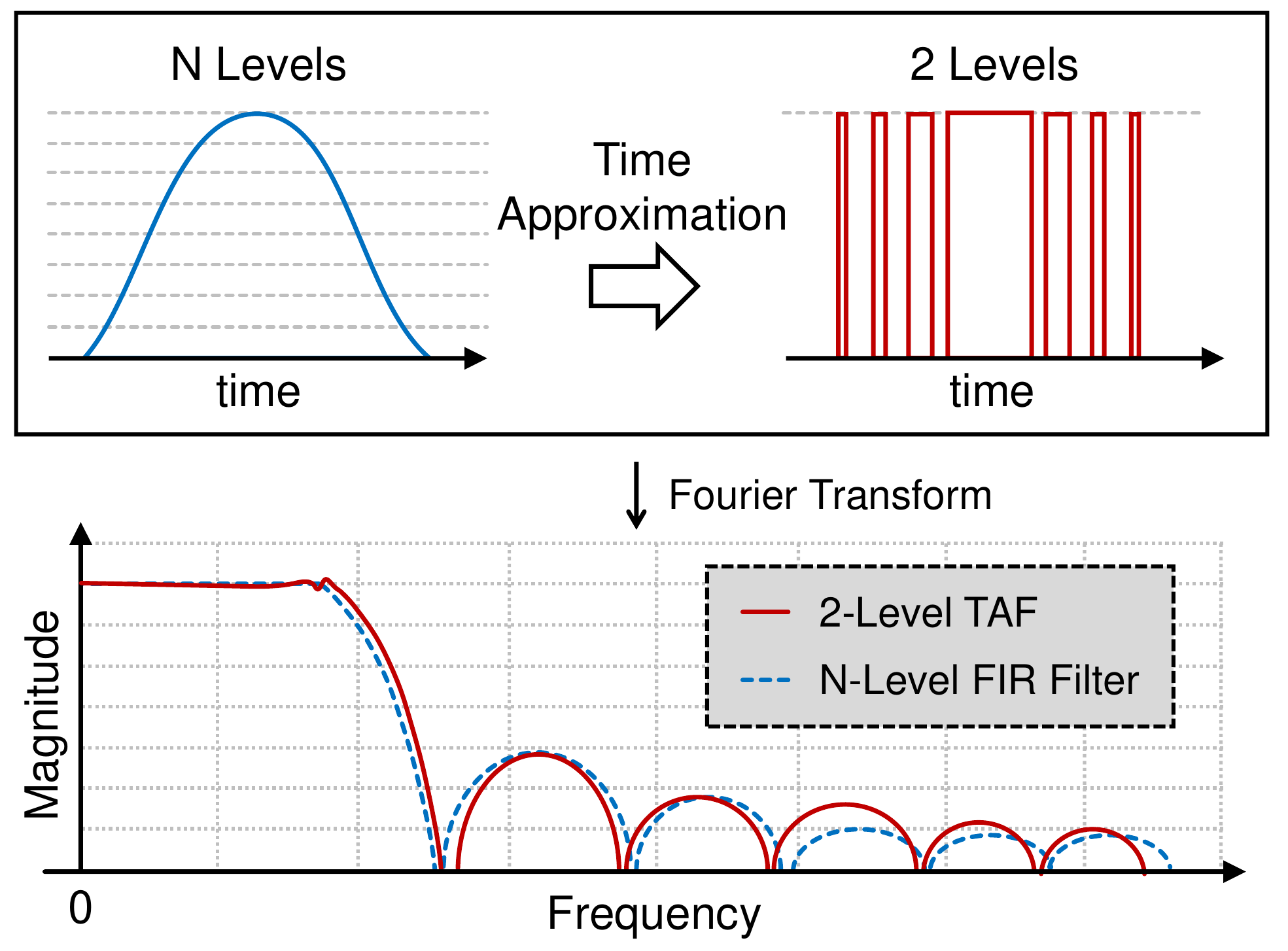}
    \caption{Time approximation scheme of AMS filter.}
    \label{proposed_TAF}
\end{figure}


The concept of time approximation scheme is shown in Fig.~\ref{proposed_TAF}. The N-level impulse response of an arbitrary AMS FIR filter is encoded into a two-level time-modulated waveform, such that these two waveforms have similar frequency responses for filtering. Conventionally, AMS FIR filter implements the filter coefficients by scaling the size of DACs. The mismatch between the DACs inevitably degrades the filter performance. In contrast, TAF only requires a single DAC to express all of the filter coefficients by multiplying the digital-like TAF pattern with the input digital signal. Therefore, there is no mismatch between the filter taps. The multiplication can be achieved via MUXs or NAND gates for a single-bit pattern with minimum extra implementation overhead. In addition, TAF allows changing the filter coefficient digitally with a wide tuning range, which makes it more flexible than the conventional AMS FIR filter.

\subsection{Time Approximation Errors}
There are mainly two types of time approximation errors for a TAF: 1) the intrinsic error of approximating an amplitude-varying response in time with a constant amplitude and 2) the time quantization error set by minimum achievable time resolution. To the first order, the intrinsic approximation error $G_{err}$ is estimated as a function of the frequency $f$,

\begin{equation}
\label{eq1}
G_{err} = 20 ~\mathrm{log_{10}} \frac{\mathrm{sinc}(a_{min}T_{tap}f)}{\mathrm{sinc}(T_{tap}f)},
\end{equation}
where $a_{min}$ is the smallest filter coefficient and $T_{tap}$ is the time interval between the adjacent filter taps. 

In reality, the resolution of time approximation is determined by the period of the fastest clock, which cannot be infinitely small. Therefore, the time quantization error due to finite time resolution degrades the TAF performance. With the presence of circuit non-idealities, TAF's response will further deviate from the original FIR filter response in terms of the locations and depth of filter notches and the in-band flatness.

\section{Proposed Filter Automation Flow}

\begin{figure}[!t]
    \centering
    \includegraphics[width=0.48\textwidth]{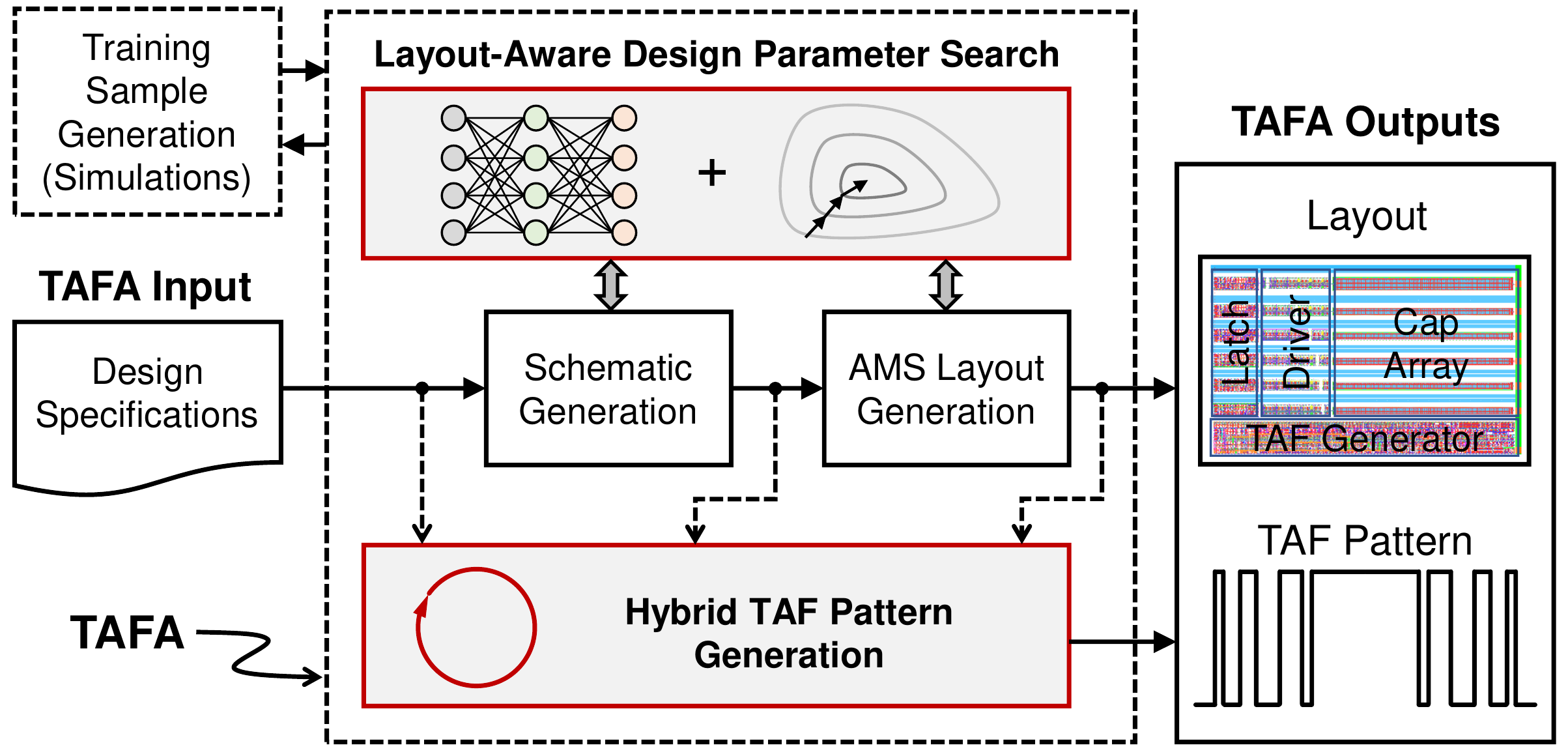}
    \caption{Proposed design automation flow for AMS filters.}
    \label{automation_flow}
\end{figure}

Figure ~\ref{automation_flow} shows the proposed AMS filter design flow. Based on the filter specifications, a hybrid TAF pattern (impulse response) generator is used to optimize the TAF pattern and minimize the time approximation errors given certain time quantization resolution. In addition, the simulated output transient waveforms of the circuit (schematic and post-layout netlists) are fed back to the TAF pattern generator for further pattern optimization considering the circuit non-idealities. To expedite the AMS filter design, we combine advanced search algorithm with a layout-aware regression model to rapidly search for the design parameters, such as device sizing and biasing. The regression model is based on an artificial neural network (ANN), which is first trained and validated using a schematic-level dataset generated from a parameterized circuit netlist and SPICE simulations \cite{AMPSE_core_2021,NN_TCAD19,zhang2020cepa}. 

To incorporate the layout parasitic information, a linear transfer learning (LTL) technique is applied to the ANN model trained using the schematic dataset \cite{liu2020tl}. The layouts are generated using the proposed mixed-signal layout automation flow, which includes the standard digital design flow and a custom place and route (P\&R) engine for analog components and top-level integration. After the layout-aware model is obtained, the design parameters are determined using a gradient based search algorithm. This layout-aware parameter search method leverages the fast speed of ANN inference and the reduced post-layout simulation overhead thanks to LTL, thus enabling efficient and low-cost optimization.

\subsection{Hybrid TAF Pattern Generation}\label{AA}

\begin{figure}[!t]
    \centering
    \includegraphics[width=0.48\textwidth]{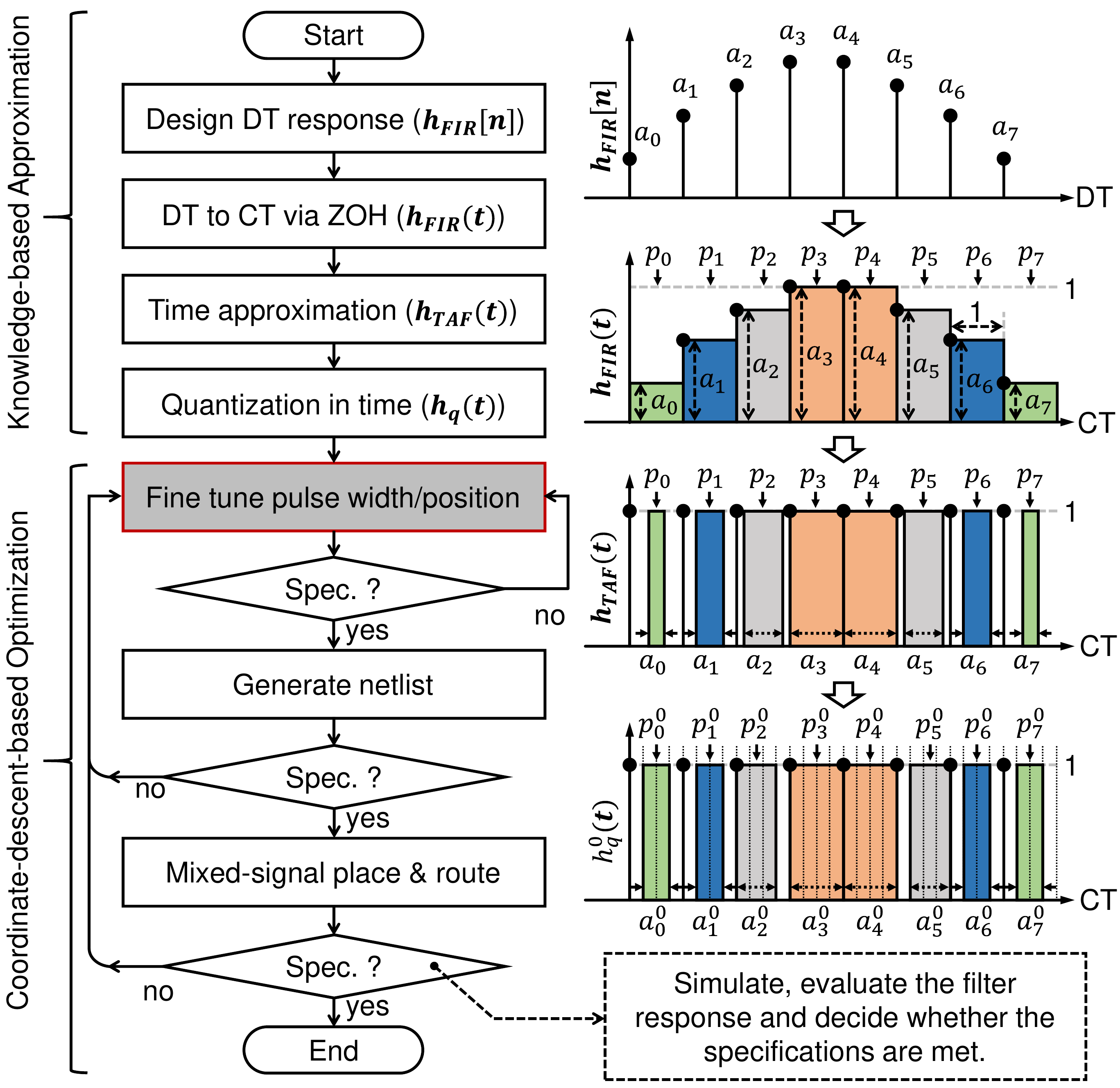}
    \caption{Proposed hybrid TAF pattern generation.}
    \label{proposed_flow}
\end{figure}

The proposed hybrid TAF pattern generation is depicted in Fig.~\ref{proposed_flow}. To avoid inefficient search in an enormous space of TAF pattern, a discrete-time (DT) FIR filter response is first designed based on the filter specifications ($h_{FIR}[n]$), and then converted into continuous-time (CT) response via zero-order hold (ZOH) interpolation ($h_{FIR}(t)$). To create TAF response ($h_{TAF}(t)$) from the CT FIR response, a time approximation on the filter coefficients is performed via pulse width ($a_n$) and position ($p_n$) modulations, such that the area and the position of each pulse is the same as that in $h_{FIR}(t)$. Pulse width and position are then quantized by the finite time resolution. The quantized time-approximated filter response ($h_{q}^0(t)$) is typically not optimal at this stage due to the time approximation errors and circuit non-linearity. An optimization loop based on coordinate descent is employed to fine tune the pulse width ($a_n^i$) and position ($p_n^i$) of each filter coefficient, so that TAF performance is maintained or improved in the presence of the aforementioned non-idealities. 

\begin{figure}[!t]
    \centering
    \includegraphics[width=0.46\textwidth]{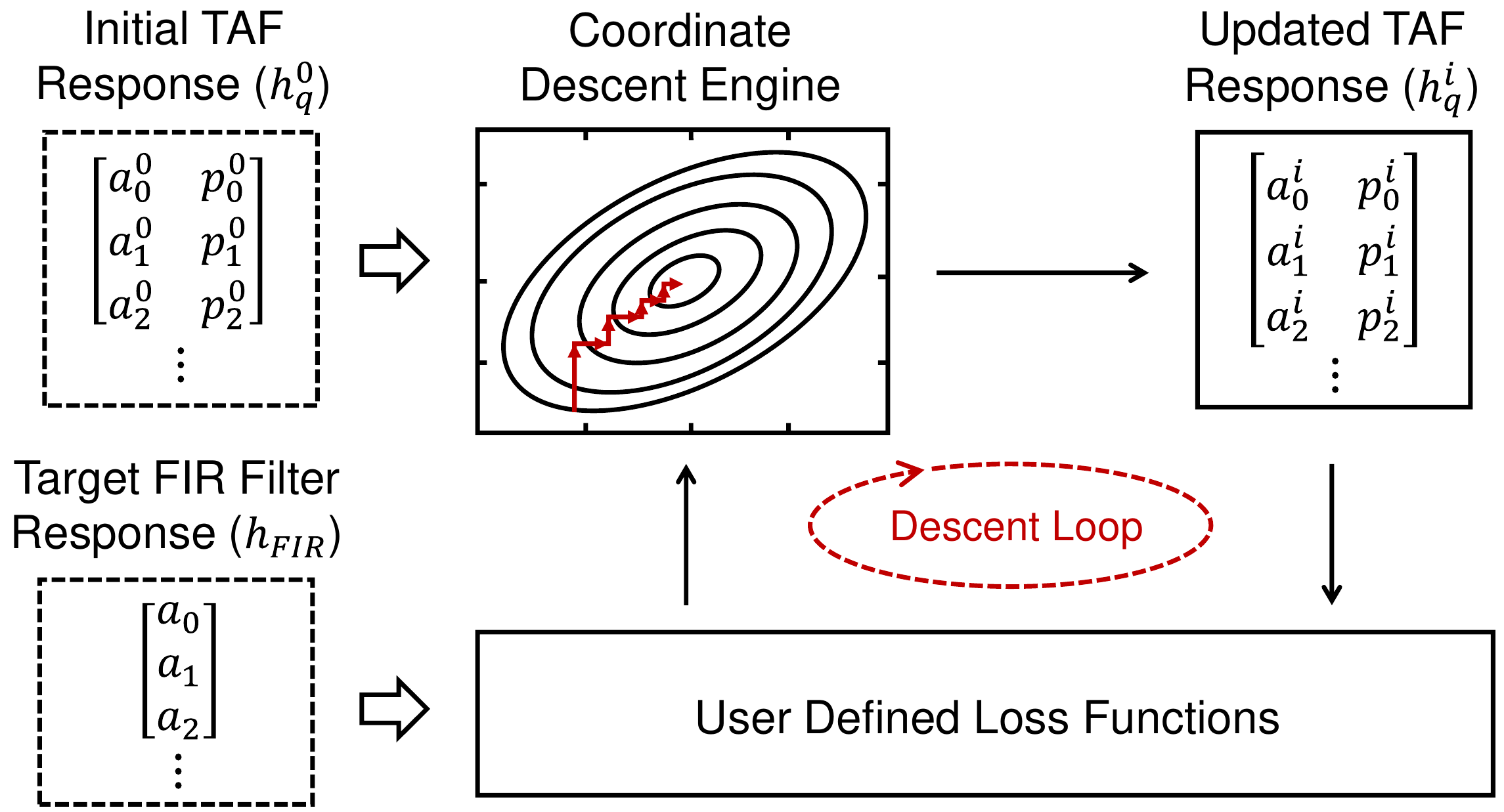}
    \caption{Fine tune scheme of TAF pattern using coordinate descent.}
    \label{fine_tune}
\end{figure}

Figure~\ref{fine_tune} presents the proposed fine tune scheme. The TAF response obtained from the knowledge-based approximation steps ($h_q^0$) is used as the initial point of the optimization loop. Based on the loss incurred in taking a step toward a certain direction, the optimization engine decides whether to stay at the current location or move toward that direction. The updated TAF response at the $i^{th}$ iteration is annotated as $h_q^i$, and is used to calculate the loss in the next decision. In order to be flexible, \textbf{different loss functions} can be applied for different applications. If the goal is to design a TAF response that mimics the response of the target CT FIR filter within the whole Nyquist band ($DC$ to $B$) as closely as possible, the loss function is designed as:

\begin{equation}
\label{eq2}
L(B)=\frac{1}{B}\int\displaylimits_{0}^{B} (|H_{FIR}(f)|-|H_q^n(f)|)df.
\end{equation}
The function utilizes the normalized magnitude difference between the frequency responses of the target FIR filter and the associated TAF. On the other hand, for applications that require extremely low noise at a specific band of interest, the loss function is designed as
\begin{equation}
\label{eq3}
L(B_1,B_2,f_0)=\int\displaylimits_{0}^{B_1}\frac{|H_q^n(f)|}{B_1}df-\int\displaylimits_{f_0}^{f_0+B_2}\frac{|H_q^n(f)|}{B_2}df,
\end{equation}
which uses the difference between the integrated energy within the signal band ($DC$ to $B_1$) and the integrated energy of the specific band of interest ($f_0$ to $f_0+B_2$). Given the fixed in-band energy, lower loss leads to deeper noise suppression at the specific band. Note that other potential loss functions can be applied for different optimization objectives under the same framework. 

\begin{figure}[!t]
    \centering
    \includegraphics[width=0.49\textwidth]{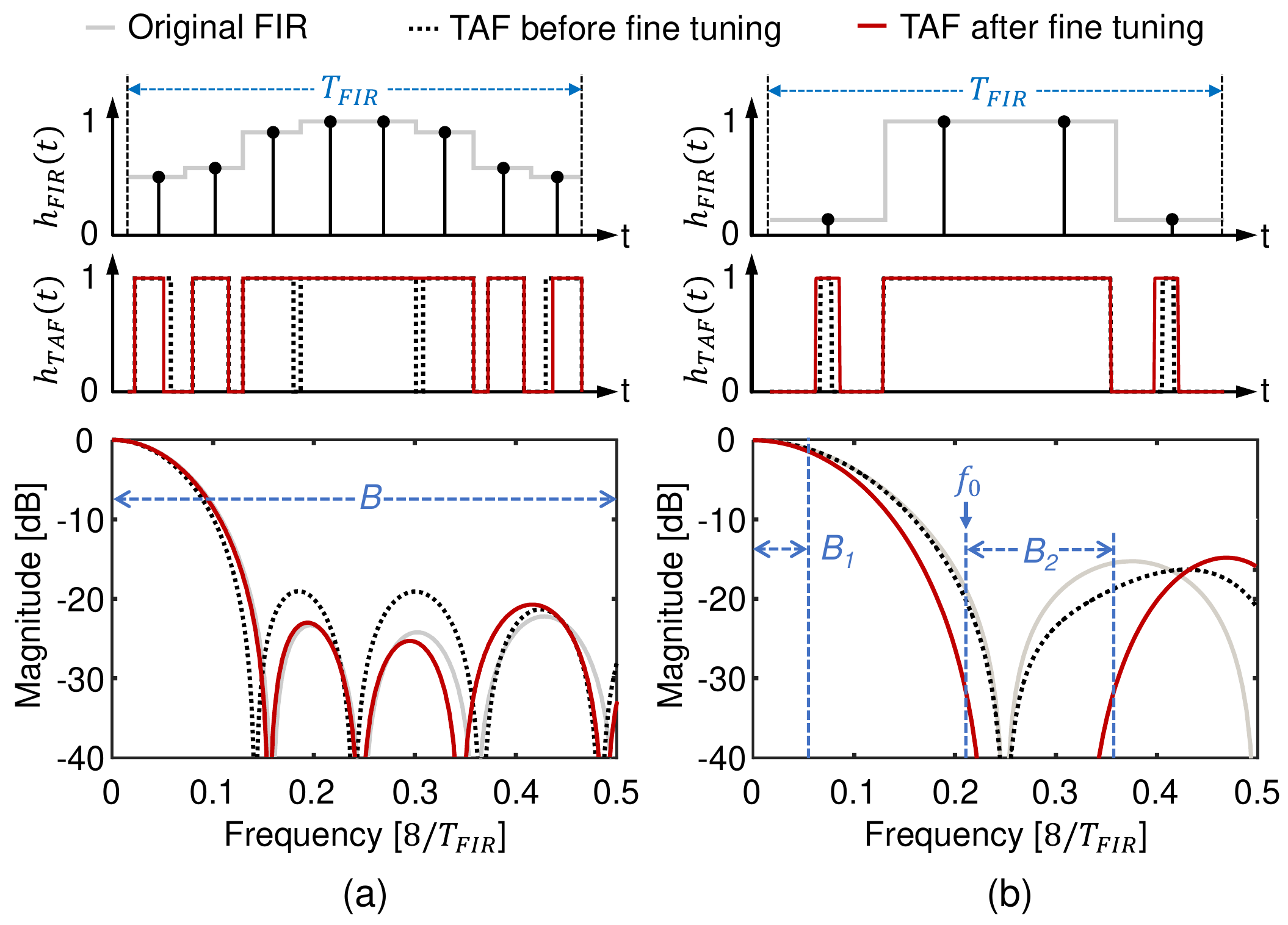}
    \caption{Filters' impulse responses and corresponding spectra of frequency responses with and without fine tuning, using (a) loss function (\ref{eq2}) and (b) loss function (\ref{eq3}), respectively.}
    \label{spectrum_fine_tuning}
\end{figure}


\begin{figure*}[!t]
    \centering
    \includegraphics[width=0.94\textwidth]{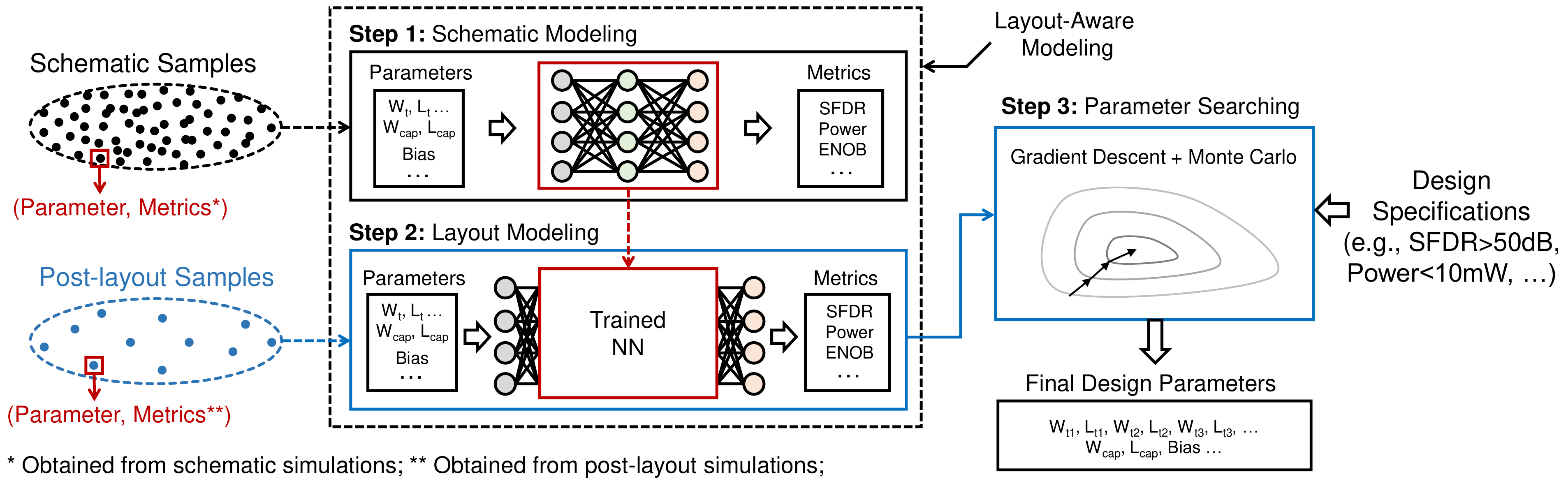}
    \caption{Proposed layout-aware AMS circuit modeling and design parameter search.}
    \label{ampse_tl_flow}
\end{figure*}

Figure~\ref{spectrum_fine_tuning} shows two examples of TAF response optimizations with different loss functions. In Fig.~\ref{spectrum_fine_tuning}(a), loss function (\ref{eq2}) is used for approximating an 8-tap FIR filter. A TAF response closer to the target FIR filter response is achieved via the proposed fine-tuning scheme. In Fig.~\ref{spectrum_fine_tuning}(b), a 4-tap FIR filter is first designed and approximated in time with the knowledge-based approach. Then, loss function (\ref{eq3}) is applied to tweak the TAF response so deeper attenuation at $B_2$ can be achieved.

\subsection{Layout-Aware Circuit Parameter Search}


The proposed layout-aware circuit parameter search is shown in Fig. \ref{ampse_tl_flow}. First, we generate an ANN model representing the schematic-level parameter-to-metric (P2M) function (\textbf{step 1} in Fig. \ref{ampse_tl_flow}). We select a large number of parameter combinations using random sampling from a predefined range of the circuit parameters and find the corresponding performance metrics via schematic-level SPICE simulations. 

Next, layout parasitic information is incorporated in the ANN model (\textbf{step 2} in Fig. \ref{ampse_tl_flow}). Obtaining training dataset (parameters-metrics pairs) in layout design stage is extremely inefficient compared to the schematic stage due to the long SPICE simulation time needed for post-layout designs. To reduce the number of required training samples for the layout model, we reuse the schematic-level ANN model and add a linear layer at its input and output. The weights and biases of the original ANN remain fixed, and only the added linear layers are trained using the post-layout simulation data. 



\begin{figure}[!t]
    \centering
    \includegraphics[width=0.44\textwidth]{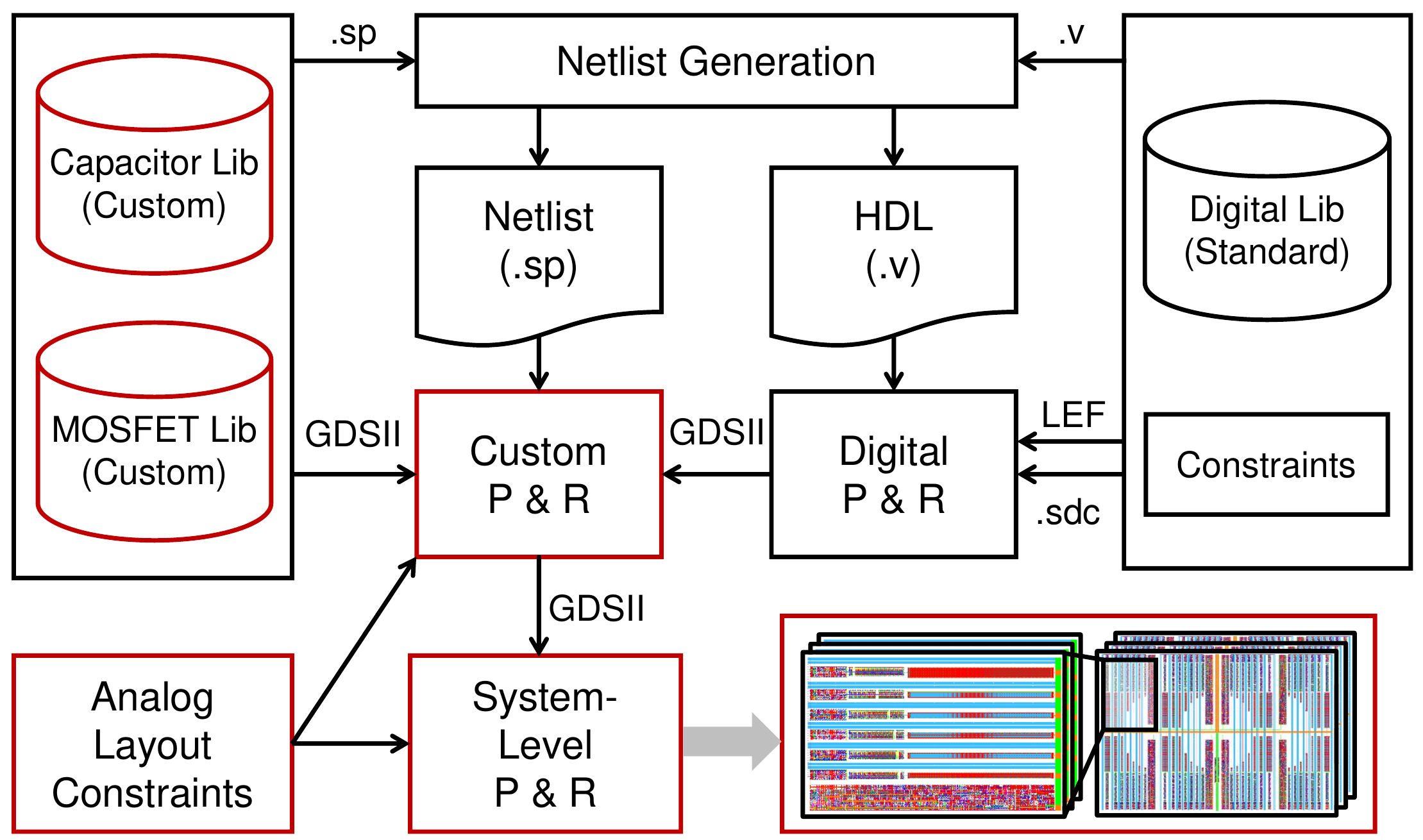}
    \caption{Mixed-signal layout automation flow.}
    \label{layout_flow}
\end{figure}

Even though the cost of training the ANN model is not low, we only need to go through steps 1 and 2 once for a specific circuit as a tool developer. After obtaining the layout-aware ANN model, as a tool user, we search for the circuit parameters (\textbf{step 3} in Fig. \ref{ampse_tl_flow}) which satisfy the desired specifications. Thanks to the quick evaluation using the ANN model, TAFA can rapidly determine the circuit parameters for diverse specifications. At each search iteration, the gradient-based algorithm changes the design parameters in a direction which minimizes a loss defined by the difference between ANN predicted metrics and the target specs. A parallel Monte Carlo optimization is also adopted to enable a fast and optimum parameter search \cite{AMPSE_core_2021}. By utilizing multiple random initial parameters, the search process can avoid getting trapped in a local minimum of the loss function. Using this technique, we can generate multiple design candidates which can satisfy the given specifications within a short time frame.

\subsection{Mixed-Signal Layout Automation Flow}

The mixed-signal layout generation is a hybrid flow that utilizes both standard digital design tools and a custom synthesis engine, as shown in Fig.~\ref{layout_flow}. Critical analog blocks including the DAC drivers and capacitors need to be carefully designed and laid out to guarantee the filter performance. Therefore, we developed a custom P\&R engine based on an open-source Python framework, taking insights from analog circuit designers. Considering the symmetry requirement for P\&R and dummy cells for load balancing, layout templates are encoded in the P\&R scripts. Similar to \cite{crossley2013bag} and \cite{ding2018circuit}, the layout templates are parameterized for different device sizes. The libraries of devices, including custom MOSFETs and capacitors, are prepared and used by the custom P\&R for different designs. Note that these analog blocks can also be laid out using other state-of-the-art P\&R tools, which leverage the advanced algorithms to generate the layouts without providing the detailed layout template by the designers, as seen in \cite{xu2019magical,kunal2019align}. Those alternatives can be seamlessly integrated into the TAFA flow. To leverage the well-developed digital design flow, we herein propose a capacitor DAC structure and a counter-based TAF waveform generator, which are more digitally-intensive implementations than the design in \cite{su2020jssc}. As a result, most of the circuits become less sensitive to mismatch and symmetry requirements and can be generated by the digital synthesis tools using a digital standard cell library. Behavioral Verilog files and timing constraints are prepared for digital synthesis and P\&R. In the end, the custom P\&R integrates the digital-synthesized layouts with the custom layouts for a single channel of the TAF. To generate the whole TAF, the system-level P\&R in the flow can integrate eight time-interleaved channels symmetrically considering the power, signal, and clock routing. Such  hierarchical layout generation is another feature of the proposed flow.

\section{Experimental Results}

\subsection{Embodiment of Filter Implementation}

\begin{figure}[!t]
    \centering
    \includegraphics[width=0.48\textwidth]{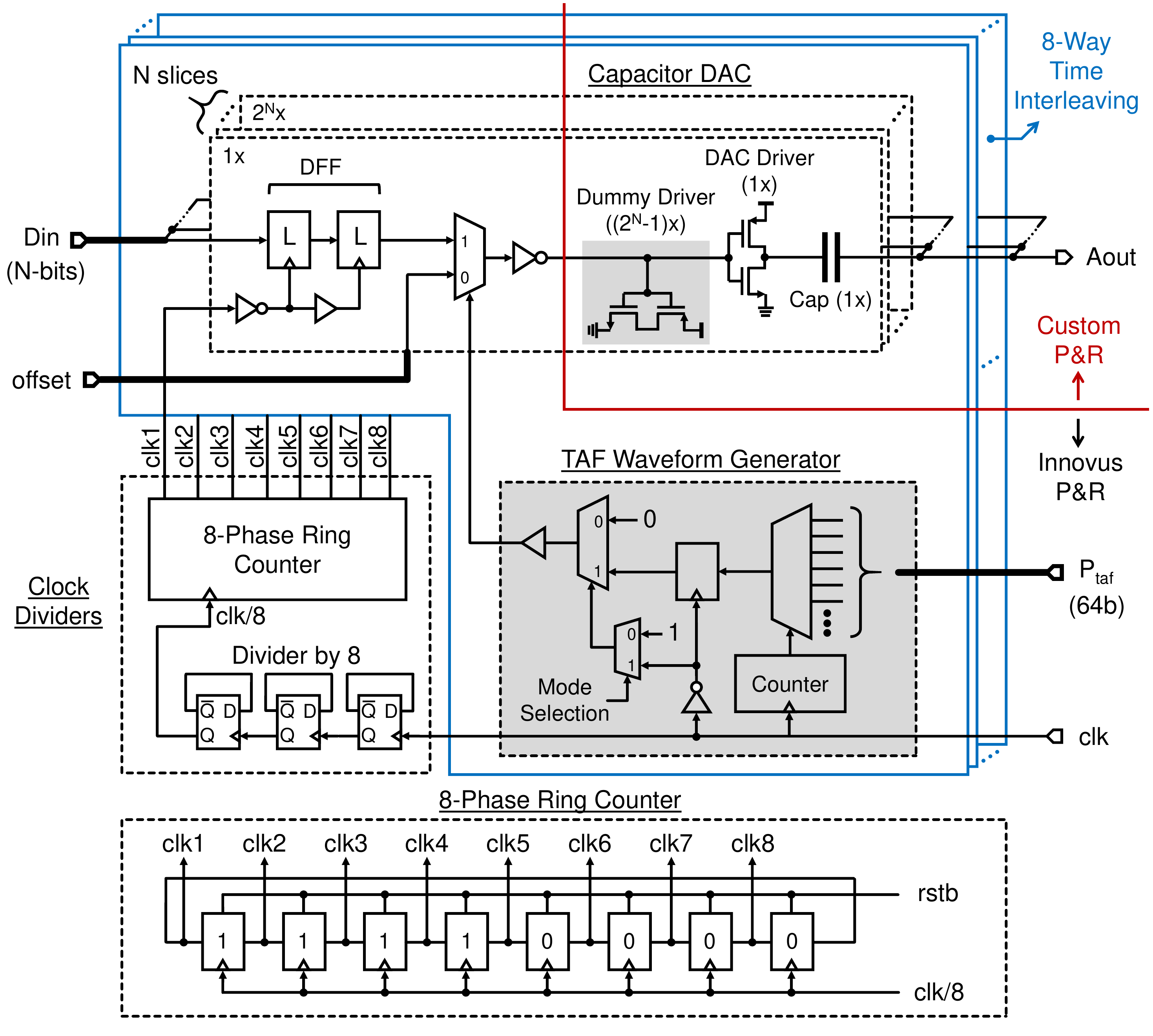}
    \caption{System block diagram of the proposed AMS filter.}
    \label{system_block}
\end{figure}

The block diagram of the TAF implementation is shown in Fig.~\ref{system_block} and is used to demonstrate the proposed TAFA flow. To extend the impulse response of the TAF, an eight-way time-interleaved (TI) structure is used. Each TI channel consists of a capacitor DAC and a counter-based TAF waveform generator. The clock signal (clk) used for the TAF waveform generation is divided eight times via the feedback flops and further divided into eight slower clocks with shifted phases for TI operation. The shifted phases are generated by a ring counter, as shown in Fig.~\ref{system_block}. The TAF pattern is synthesized with the proposed hybrid flow and stored as static digital signals. The TAF waveform generator uses a MUX and a counter to serialize these static signals into a 1-bit data stream, which is then used to modulate the DAC input signal for the TAF. In the circuit example, the TAF pattern contains 64 bits. A DFF is inserted at the output of the MUX to avoid glitches of the MUXed signal and to simultaneously force the timing resolution of the TAF to be one period of the clk signal, which guarantees zero timing skew between taps of the TAF. The TAF waveform generator is able to synthesize both lowpass and bandpass filter responses by passing or chopping the MUXed signal, respectively.


\begin{figure}[!t]
    \centering
    \includegraphics[width=0.48\textwidth]{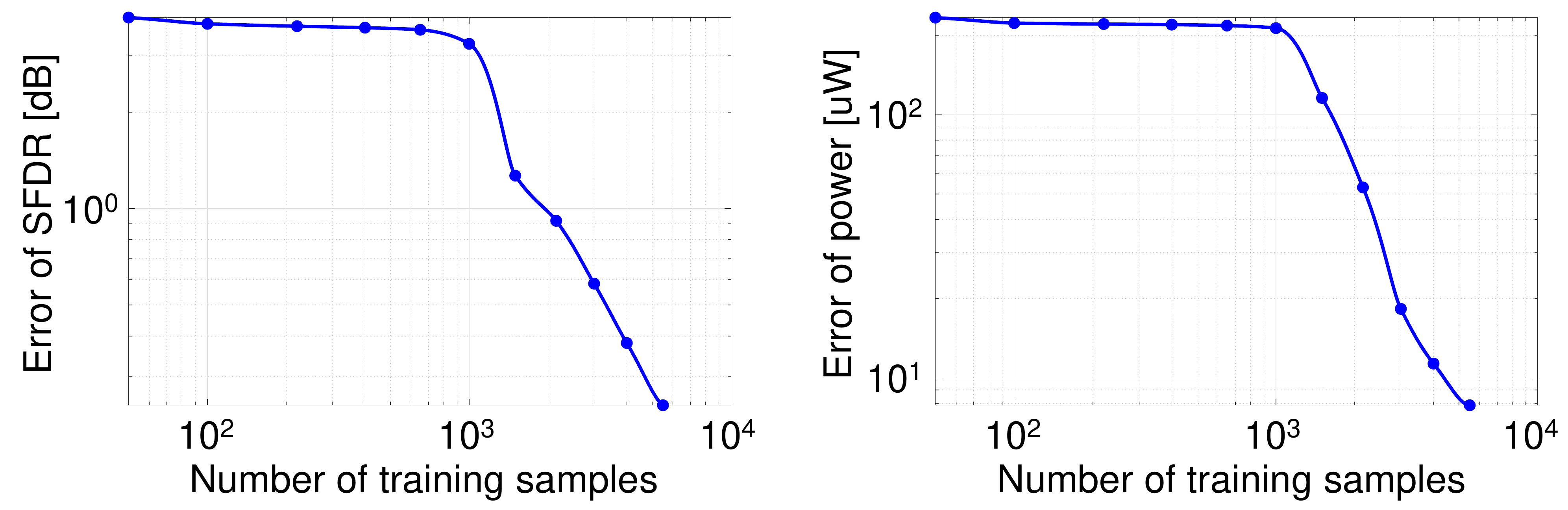}
\caption{ Testing error versus number of training samples (schematic).}
\label{err_vs_samples_sch}
\end{figure}

\begin{figure}[!t]
    \centering
    \includegraphics[width=0.48\textwidth]{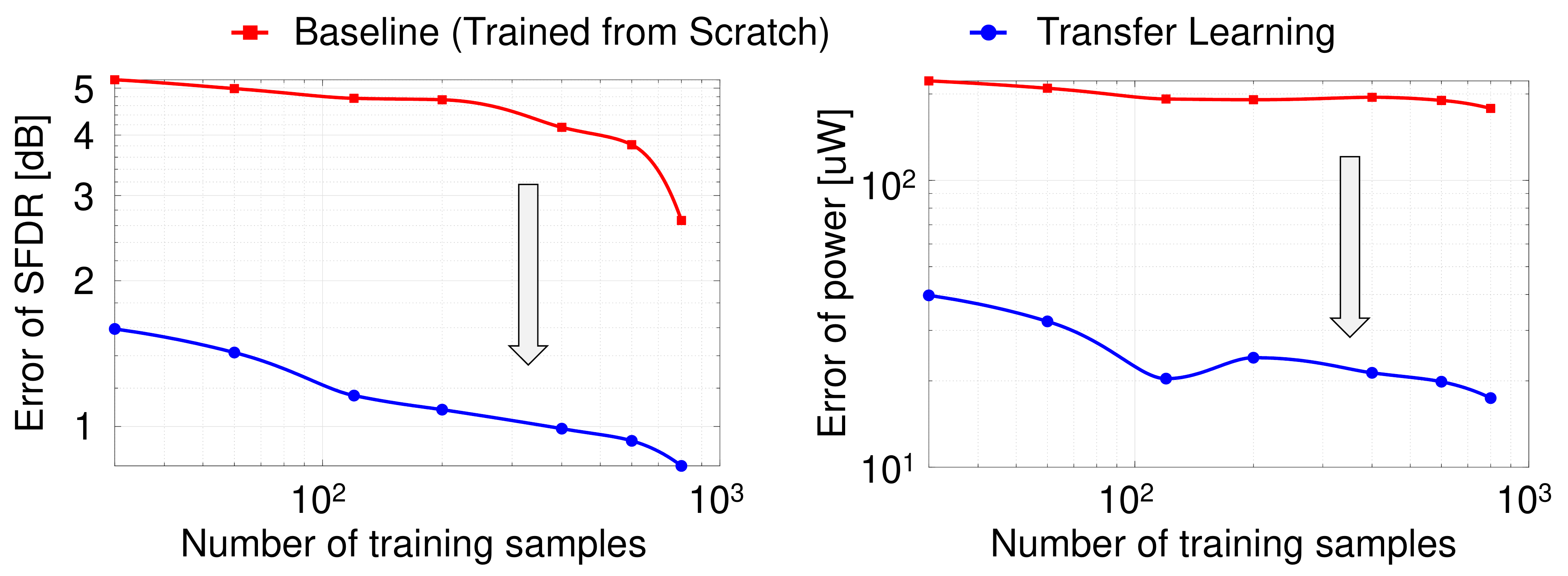}
    \caption{ Testing error versus number of training samples (layout).}
    \label{err_vs_samples_tl}
\end{figure}

\subsection{Preliminaries and Circuit Synthesis}


The AMS filter uses 10 different circuit parameters and two performance metrics which are power and SFDR. The associated ANN model has 3 hidden fully-connected layers with 128, 256 and 128 neurons, respectively. In principle, the hyper-parameters (ANN structure, training epoch etc.) are chosen to minimize the testing loss and prevent over-fitting. We randomly sample the parameter space to generate the training dataset for the ANN. As shown in  Fig.~\ref{err_vs_samples_sch}, the surrogate model can accurately predict the circuit performance when sufficient training samples are guaranteed. In this example, we use 5,500 training samples from schematic simulations. 

Next, LTL is applied to generate the post-layout P2M model of the circuit. We generate designs via random sampling, employ the proposed mixed-signal layout flow to obtain the corresponding layouts, and use the post-layout simulation results as metrics to train the layout-aware ANN model. With a less number of training samples (i.e. 800 samples from post-layout simulations), LTL achieves a much lower testing error compared to an ANN that is trained from scratch, as shown in Fig.~\ref{err_vs_samples_tl}. Note that the ANN model is only prepared once and then used for designing various AMS filters.

\begin{table}[!t]
    \caption{Performance Summary}
    \centering
    \includegraphics[width=0.44\textwidth]{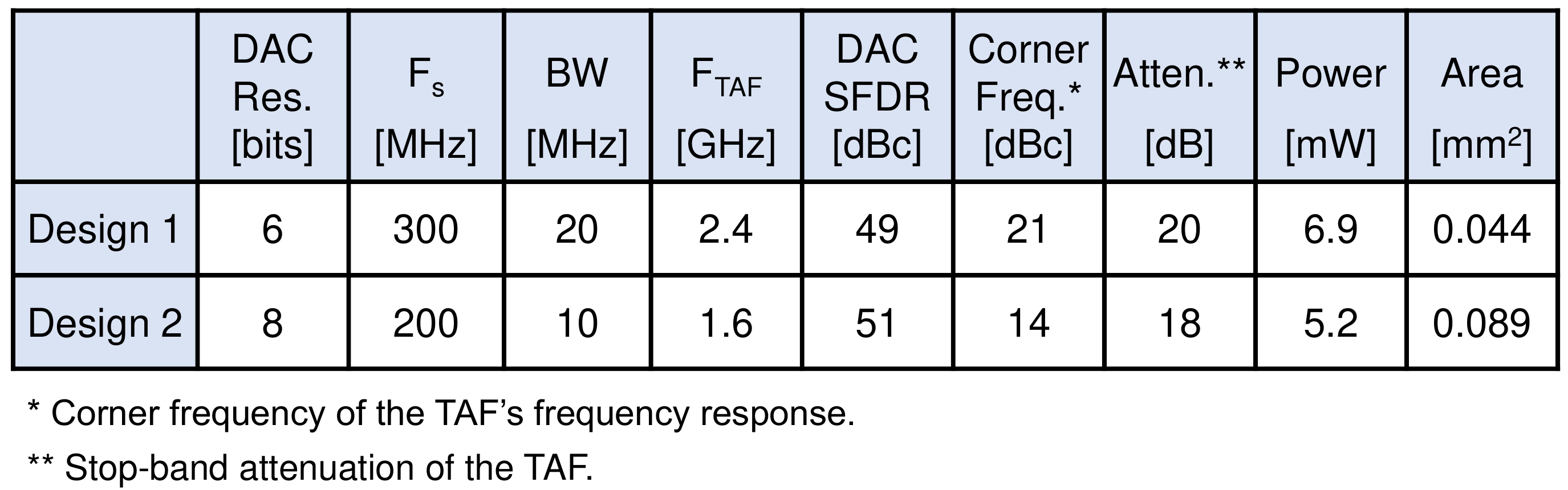}
    \label{performance_summary}
\end{table}

\begin{figure}
    \centering
    \includegraphics[width=0.44\textwidth]{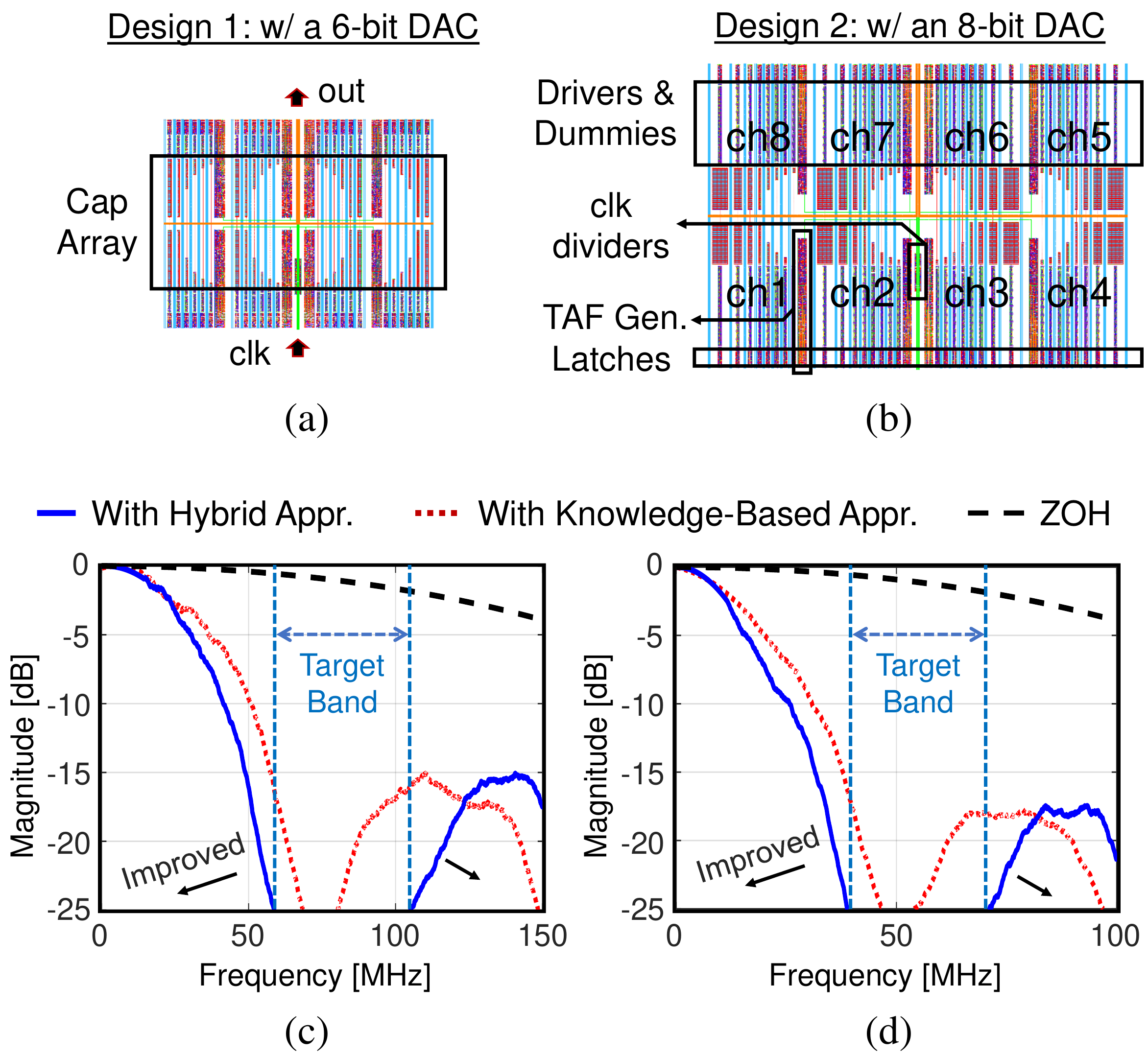}
    \caption{Layouts of (a) design 1 and (b) design 2; simulated transfer functions of (c) design 1 and (d) design 2.}
    \label{layout_tf}
\end{figure}

Based on this layout-aware ANN model, the circuit sizing is automated using the gradient-based search algorithm in combination with parallel Monte Carlo. For 10 different target specifications, this experiment takes less than 34 seconds to finish the parameter search with a six-core 2.60GHz CPU and a 32GB RAM. Given the design parameter, the proposed layout generator takes around 30s to generate the layout of the single-channel TAF, and around 10s to finish the P\&R for the eight-way TI TAF. 

\subsection{SPICE Validation of the Synthesized AMS Filter}

\begin{table}[!t]
    \caption{Performance Comparison}
    \centering
    \includegraphics[width=0.44\textwidth]{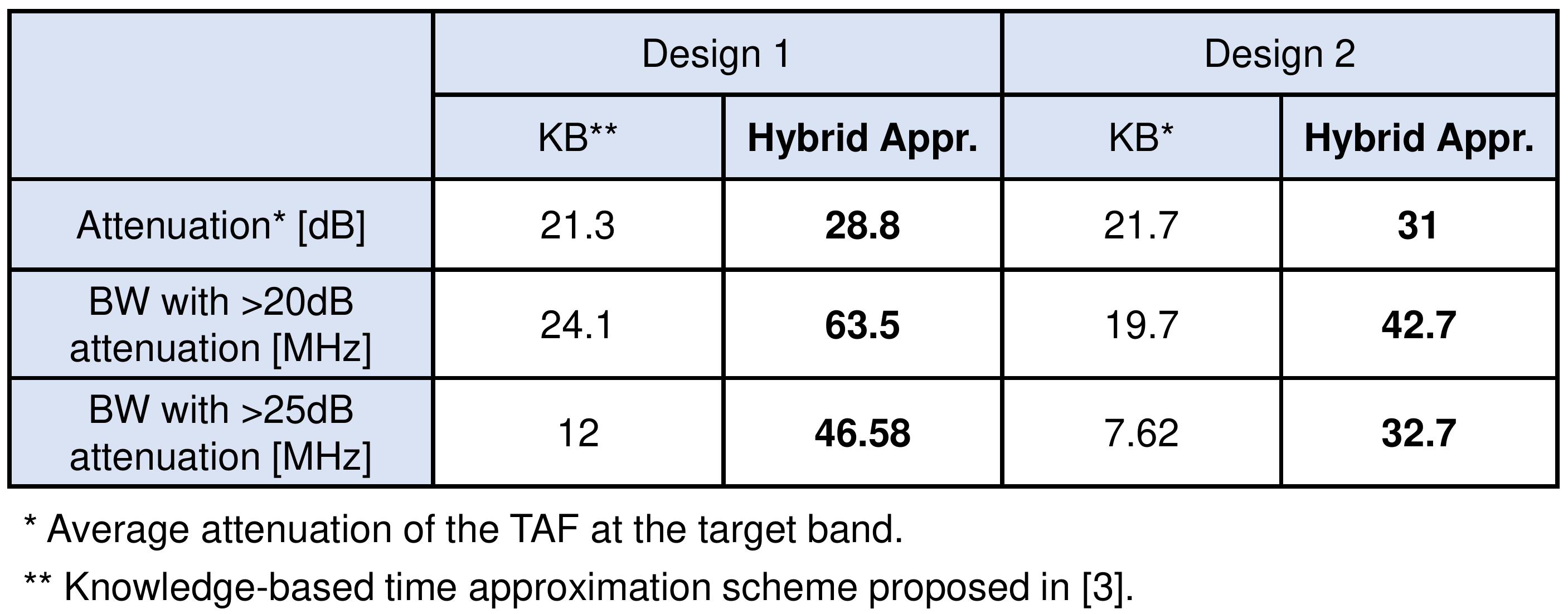}
    \label{comparison_KB_hybrid}
\end{table}

The performance of two representative TAF designs in 65nm CMOS process with different specifications is summarized in TABLE~\ref{performance_summary}. Each design is selected from 10 candidates that meet the design specifications. The circuits are designed and laid out automatically via the aforementioned automation techniques. The results are obtained via post-layout simulations with the SPICE model. With the proposed mixed-signal layout flow, the TAF waveform generator is able to operate faster than 2.4GHz ($\mathrm{F_{TAF}}$). Decent DAC linearity and stop-band attenuation are also achieved given the power and area budgets.

\begin{figure}
    \centering
    \includegraphics[width=0.48\textwidth]{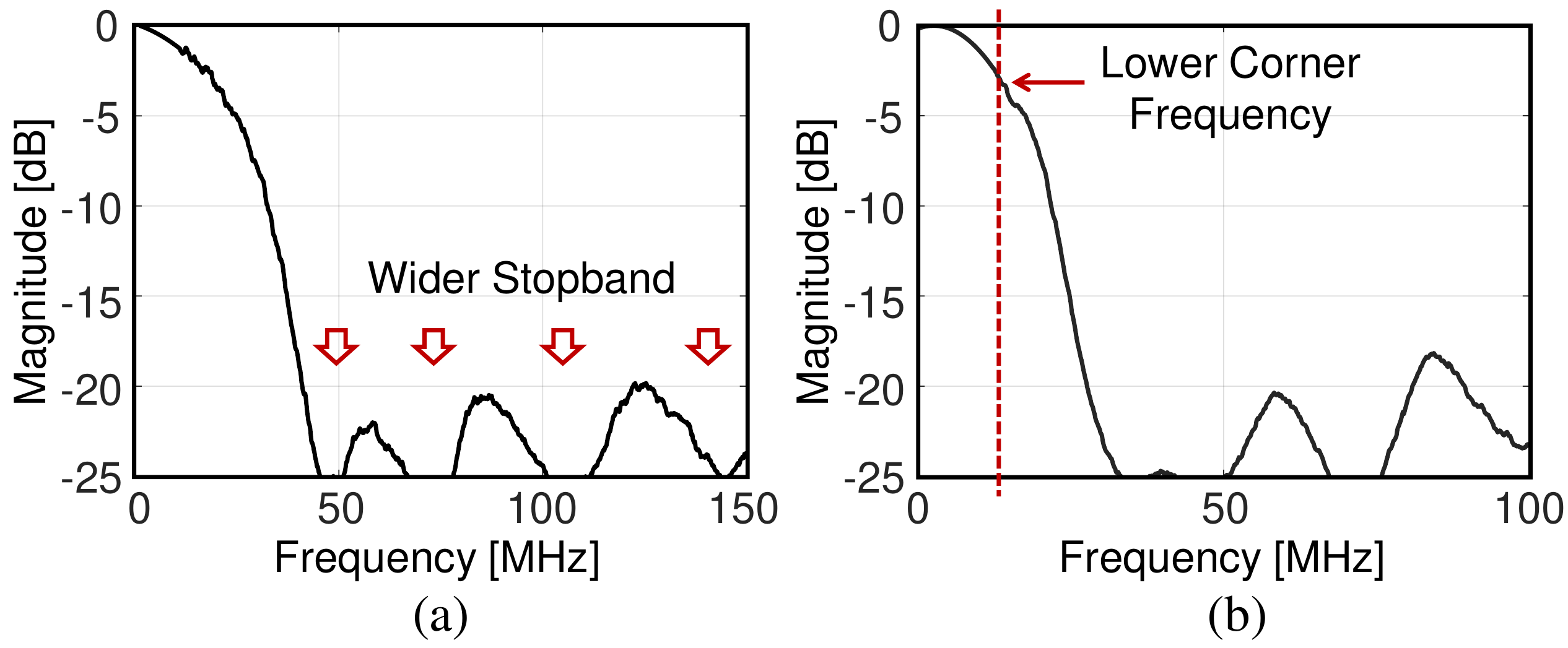}
    \caption{Simulated transfer functions of (a) design 1 and (b) design 2.}
    \label{spectrum_narrow_BW}
\end{figure}

\begin{figure}[!t]
    \centering
    \includegraphics[width=0.48\textwidth]{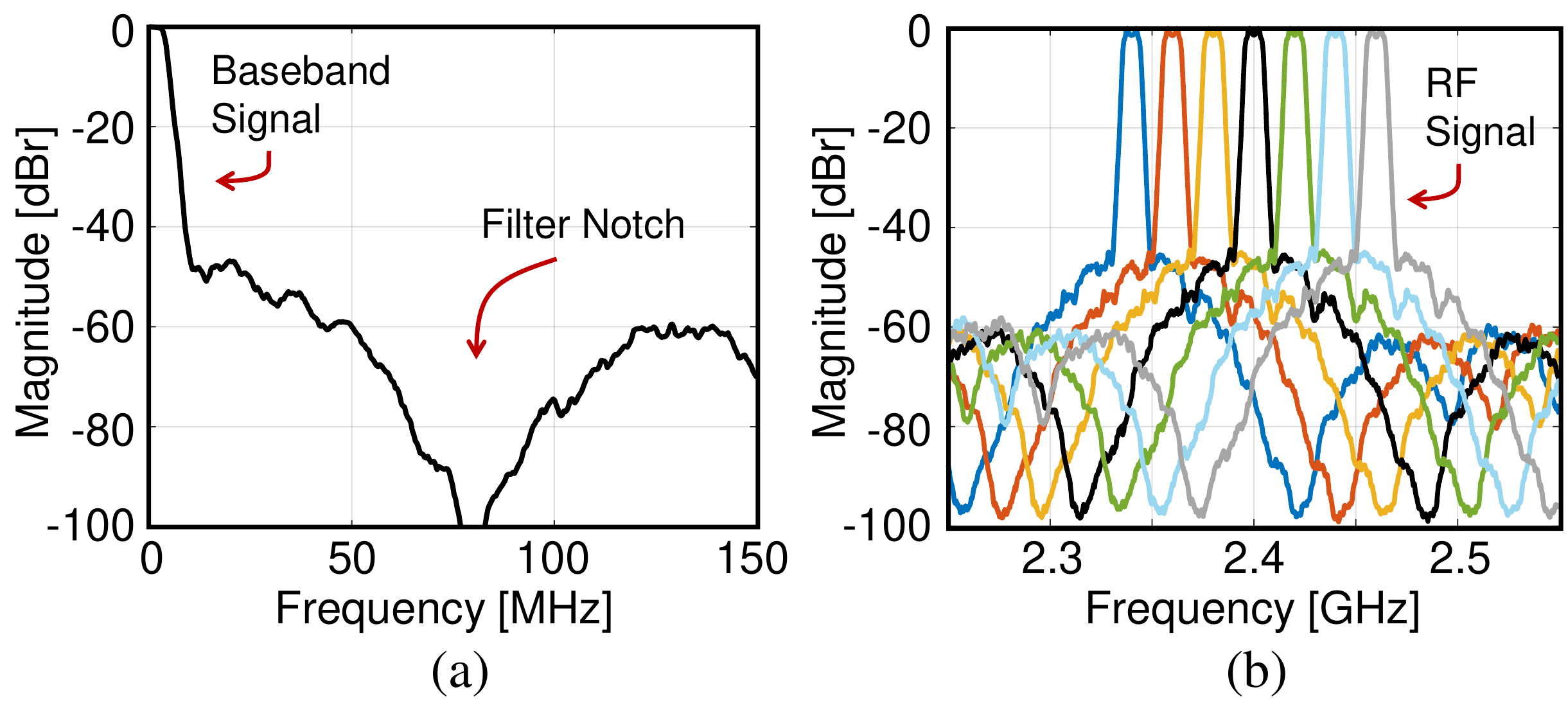}
    \caption{Spectra of (a) a lowpass and (b) a bandpass TAF output.}
    \label{spectrum_qam}
\end{figure}

Fig.~\ref{layout_tf}(a)-(b) show the generated layout for the two designs. The simulated filter transfer functions are shown in Fig.~\ref{layout_tf}(c)-(d). The filter herein aims to suppress the noise at a specific band of interest (target band) with a deep attenuation. A target CT FIR filter response is first designed based on the filter specification, and then loss function (\ref{eq3}) is applied to maximize the difference of integrated energy between the signal band and the target band. The TAF performance comparison between the knowledge-based \cite{su2020jssc} and the proposed hybrid time-approximation schemes is summarized in  TABLE~\ref{comparison_KB_hybrid}. The proposed technique shows more than 7.5dB improvement on noise attenuation and 2.2X to 4.3X bandwidth enhancement given the attenuation specifications. To prove the flexibility of the proposed AMS filter, we use TAFA to synthesize another filter response with lower corner frequency and wider stopband attenuation. The simulated transfer functions of the filter are shown in Fig.~\ref{spectrum_narrow_BW}. The filter's impulse responses are designed to generally filter the out-of-band noise with decent attenuation. In this case, loss function (\ref{eq2}) is used to reduce the magnitude difference between the frequency responses of the target AMS filter and the TAF. To validate the bandpass mode or mixing mode of the filter in the context of a wireless transmitter, we applied a 10-MHz modulated signal at 256 QAM as the input of the filter. The simulated spectra of the modulated signal at the baseband and RF band are shown in Fig.~\ref{spectrum_qam}. Deep noise reduction at the desired bands is achieved via lowpass and bandpass filtering, thus proving the dual-mode operation capability of the filter. Different carrier frequencies are swept around 2.4GHz, which shows the flexibility of the system thanks to the mostly digital architecture.

\section{Conclusion}
To address the increasing demand of AMS filter and its design challenges, this paper presented a new way of designing AMS FIR filter, featuring a time approximation architecture with associated hybrid optimization scheme for filter's impulse response and an AMS circuit design automation flow that combines a layout-aware ANN model and a gradient-based search algorithm. The fully synthesized AMS filters achieved high stopband attenuation and the proposed framework demonstrated flexibility with various filter specifications.

\section*{Acknowledgment}
The work is supported in part by DARPA ERI POSH program under Grant FA8650-18-2-7853 and in part by GlobalFoundries. The authors would like to thank Prof. Anthony F. J. Levi and Prof. Sandeep K. Gupta from the University of Southern California for technical discussions. 

\input{mybib.bbl}

\bibliographystyle{IEEEtran}
\bibliography{mybib}

\end{document}